# Transient Grating Spectroscopy of Photocarrier Dynamics in Semiconducting Polymer Thin Films

**Wenkai Ouyang**[a*], **Yu Li**[a*], **Brett Yurash**[b*], **Nora Schopp**[b], **Alejandro Vega-Flick**[a], **Viktor Brus**[bc†], **Thuc-Quyen Nguyen**[b‡] **and Bolin Liao**[a§]

[a]Department of Mechanical Engineering, University of California, Santa Barbara, CA 93106, USA

[b]Department of Chemistry and Biochemistry, University of California, Santa Barbara, CA 93106, USA

[c]Department of Physics, School of Sciences and Humanities, Nazarbayev University, Nur-Sultan City, 010000, Republic of Kazakhstan

[*] These authors contributed equally.
[†] viktor.brus@nu.edu.kz
[‡] quyen@chem.ucsb.edu
[§] bliao@ucsb.edu

**Abstract**

While charge carrier dynamics and thermal management are both keys to the operational efficiency and stability for energy-related devices, experimental techniques that can simultaneously characterize both properties are still lacking. In this paper, we use laser-induced transient grating (TG) spectroscopy to characterize thin films of the archetypal organic semiconductor regioregular poly(3-hexylthiophene) (P3HT) and its blends with the electron acceptor [6,6]-phenyl-C61-butyric acid methyl ester (PCBM) on glass substrates. While the thermal response is determined to be dominated by the substrates, we show that the recombination dynamics of photocarriers in the organic semiconductor thin films occur on a similar timescale and can be separated from the thermal response. Our measurements indicate that the photocarrier dynamics are determined by multiple recombination processes and our extracted recombination rates are in good agreement with previous reports using other techniques. We further apply TG spectroscopy to characterize another conjugated polymer and a molecular fluorescent material to demonstrate its general applicability. Our study indicates the potential of transient grating spectroscopy to simultaneously characterize thermal transport and photocarrier dynamics in organic optoelectronic devices.

A thorough understanding of the dynamics of the photogenerated charge carriers in semiconducting polymers is critical for improving organic photovoltaics (OPVs) and optoelectronics[5]. Several experimental techniques have been developed to probe the photocarrier dynamics in semiconducting polymers, targeting different time windows after photoexcitation. For example, ultrafast optical spectroscopy[6,7] typically examines the sub-nanosecond dynamics, where photogenerated excitons play a significant role. Time-resolved microwave conductivity (TRMC) measurements can resolve the dynamics of free charge carriers in longer time scales, from hundreds of nanoseconds to microseconds[8,9]. Transport and recombination of dissociated free charge carriers (polarons) in these longer time scales directly influence OPV devices' performance. Also, device-level current-voltage (I-V) characterization[10] can provide indirect information about the photocarrier dynamics at the macroscopic scale. Given the complexity of photocarrier dynamics in semiconducting polymers spanning wide space and time scales, it is desirable to have a time-resolved technique that can examine both transport and recombination of photocarriers with conveniently adjustable length scales. Moreover, as effective heat dissipation becomes increasingly important for device efficiency[11], simultaneous characterization of photocarrier dynamics and thermal transport properties of semiconducting polymers becomes an attractive feature. So far, several experimental techniques have been demonstrated to probe thermal conduction in organic materials, including time-domain thermoreflectance (TDTR)[12] and using suspended microdevices[13]. Though these transient and steady-state methods have been employed for measurements of thermal properties in bulk and thin film materials[14], they are not capable of monitoring charge carrier dynamics. It is known that the properties of polymeric materials are subjected to variations due to processing and environmental conditions, further strengthening the demand to characterize both carrier dynamics and thermal transport in a single measurement.

Laser-induced transient grating (TG) spectroscopy is an optical technique successfully used in a wide range of applications[15], including quasiparticle dynamics, thermal transport and mass diffusion. Recently it has been demonstrated for studying heat transfer in suspended polyethylene thin films[16]. This versatile pump-probe type method is based on two interfering pulsed laser beams derived from the same laser source, which create a spatially periodic excitation profile, resulting in a transient "grating" due to a rise of local photocarrier population and/or temperature after photogenerated carriers transfer energy to the lattice and cool down[15,17–19]. The increased excited carrier population and temperature rise induce changes in the complex optical constants, which subsequently decay in time due to carrier/thermal diffusion and carrier recombination, leading to time-dependent diffraction of a probe beam. Accordingly, it is sensitive to both carrier dynamics and thermal transport in the time domain from nanoseconds to microseconds over micrometer distances that can be conveniently adjusted by changing the grating period. In addition to this advantage, TG spectroscopy can conveniently separate the contributions from transport and local recombination and dissipation by continuously controlling the transport length scale set by the grating period.

In this work, we apply the TG technique to measure the grating decays in the model system regioregular poly(3-hexylthiophene) (P3HT) and its blend with the electron acceptor [6,6]-phenyl-$C_{61}$-butyric acid methyl ester (PCBM). Charge transport in this system has been extensively studied[8,9,20–22], and a limited amount of work has been directed towards understanding the thermal properties[23–25]. In our study, we examined neat P3HT films and associated blends containing PCBM with a weight ratio of 1.8% and 50%, hereafter 1.8% and 50% blend, respectively. By varying the excitation grating period, we were able to resolve both the excited state and heat relaxation processes with the application of an appropriate model containing the two processes. It was found that the carrier dynamics are determined by multiple recombination mechanisms, while heat transfer was dominated by the response of the substrate. Our study suggests the TG technique can be a potentially versatile platform to characterize both photocarrier dynamics and thermal transport properties of semiconducting polymers.

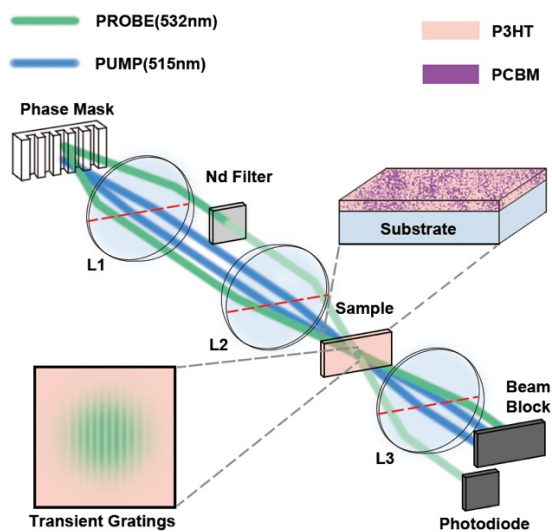

**Figure 1**. Schematics of TG measurements of P3HT:PCBM thin films on glass substrates.

In the TG setup, illustrated schematically in Fig. 1, an optical diffraction grating ("phase mask") separates the pump and probe beams into ±1 diffraction orders, and a two-lens confocal imaging system is employed to recombine the pulses at the sample at a controlled crossing angle $\theta$. The two crossed pump beams with the crossing angle $\theta$ creates an interference pattern with a period of *L = λ/[2sin(θ/2)]*, where $\lambda$ is the wavelength of the pump light. The decay of the intensity of the transient gratings is monitored by diffraction of a probe laser beam, and the grating period will determine the transport distance of both heat and photocarriers, which, in our setup, was varied between 2-10 μm. The TG setup employs optical heterodyne detection.[15,26] One arm of the probe beam is attenuated with a neutral density filter to serve as a local oscillator (the reference beam). The diffracted signal is coherently superposed with the reference beam, and then they are directed into a fast photodiode (Hamamatsu C5658). The signal is subsequently read out using an oscilloscope (Tektronix TDS784A). The heterodyne detection can increase the signal-to-noise (S/N) ratio, as well as yield a signal that is linear with respect to the material response, simplifying the analysis and interpretation of the measurements[15]. The pump pulses of 515 nm and 260 fs pulse width (Clark-MXR IMPULSE) have a fluence of 120 nJ/cm$^2$ with a diameter of 99 um and a repetition rate of 250 kHz. The continuous wave (CW) 532 nm probe beam has a power of 421 mW/cm$^2$ with a diameter of 83 μm. The time resolution of this TG implementation is limited by the bandwidths of the photodetector and the oscilloscope and estimated to be a few nanoseconds.

Samples were prepared by spin-casting solutions of P3HT:PCBM dissolved in 1,2-dichlorobenzene onto cleaned glass substrates, inside an inert atmosphere glovebox. The solution was approximately 4% wt added solids. A spin-speed of approximately 800 rpm produced samples with thickness ranging from 50 to 250 nm (determined after all measurements using an Ambios XP-100 profilometer). After spin-casting, samples were annealed in the glovebox at 150 degrees Celsius for 15 minutes, followed by encapsulation with another clean glass slide with epoxy around the edges, with a gap between the sample surface and the encapsulation slide. Encapsulation of the samples ensured no contamination by water or oxygen during the TG measurements.

Grating periods from ~3 μm to ~9 μm were adopted in our experiment, and the measured real-time decay data of the transient gratings with two periods (L~ 8.81 μm and 3.75 μm) are shown in Fig. 2. The availability of a wide range of grating periods, which set the transport length scale, is one unique advantage of the TG technique: the grating-period-dependent decay rate provides information to the diffusion coefficients, while the grating-period-independent decay rate yields local dissipation or carrier recombination[27]. The TG technique monitors the decay of the magnitude of the transient gratings in time by measuring the intensity of the diffracted probe beam. The "magnitude" of the transient gratings is the change of the local optical refractive index caused by the pump excitation. Both photocarrier generation and temperature rise can lead to the change of the optical refractive index. Thus, in principle, the TG technique is sensitive to both the heat diffusion process and the photocarrier dynamics. In inorganic materials, these two processes typically happen over distinct time scales: the photocarrier dynamics typically occur within one nanosecond[28,29] while the heat diffusion occurs in tens to hundreds of nanoseconds given micrometer-scale grating periods[19]. Therefore, this timescale separation usually prohibits simultaneous evaluation of the two processes in one measurement. In contrast, in semiconducting polymers, the photocarrier dynamics can persist for a much longer time, up to microsecond time scales, as revealed by TRMC measurements[8], which indicates that the TG signal will be potentially influenced by both photocarrier dynamics and the heat diffusion process within the nanosecond to microsecond time window, and both processes need to be considered when interpreting the TG results.

The heat diffusion process is governed by the heat equation with a 1D-periodic initial temperature distribution. Generally, TG spectroscopy is used to detect thermal transport properties of suspended films[28,30] or thick films on substrates[15,19]. In our samples consisting of absorbing thin films on transparent substrates, the heat diffusion process is dominated by the glass substrate due to the much smaller thickness of the thin films (50 to 250 nm) than the thermal penetration depth $d = L/\pi$[17], which is confirmed by finite element simulations using COMSOL. For the smallest grating period we used, the thermal penetration depth was about 1 $\mu$m, indicating that the thermal transport properties can be effectively extracted for thicker polymer films with micrometer-scale thickness. Alternatively, pushing the grating period down to the diffraction limit using large aperture optics[16] or using short-wavelength ultraviolet light[31] can also enable significant thermal contribution from the top film. In our current study, the absorbing thin films act as surface heating sources and the solution of the time dependent decay of the surface temperature profile has the form[15,17]:

$$T(t) = D \, \text{erfc}(q\sqrt{\alpha_{\text{th}} t}), \quad (1)$$

where $\text{erfc}(x) = (2\pi)^{-1/2} \int_x^\infty e^{-t^2} dt$ is the complementary error function, $q = 2\pi/L$ is the grating wave vector, $\alpha_{\text{th}}$ is the thermal diffusivity of the glass substrate, and $D$ is a fitting parameter related to the initial temperature rise. As shown in Fig. 2, there are significant deviations between the best fits using Eq. 1 and the experimental results indicating that the heat diffusion process alone cannot adequately explain the experimental results. The contribution from photocarrier dynamics must be taken into account.

In P3HT and its fullerene blends, it is well established that the photocarriers initially exist in the form of excitons, which quickly decay within a few nanoseconds in neat films and even faster in blends where the donor-acceptor interfaces strongly facilitate exciton dissociation into free charge carriers. These free charge carriers can persist much longer, up to microseconds, and in fact, their transport and recombination dynamics directly impact the performance of organic optoelectronic and photovoltaic devices[5,32]. Even in neat P3HT without electron acceptors, it is reported that a significant amount of free charge carriers (up to 30% of the total generated species) can be directly photoexcited[8,33,34]. The concentration of free charge carriers is expected to be much higher in the blends due to efficient exciton dissociation. Given the measurement window of our TG setup (~ns to ~$\mu s$), the contribution to our TG signal is most likely from the dynamics of the photogenerated free charge carriers. Considering first-order and second-order recombination processes[8,20,35] and carrier diffusion, the spatial-temporal evolution of the free charge carrier concentration $n(x,t)$ obeys the governing equation:

$$\frac{\partial n}{\partial t} = \alpha_{pc} \frac{\partial^2 n}{\partial x^2} - \frac{n}{\tau} - Bn^2, \qquad (2)$$

where $\alpha_{pc}$ is the diffusivity of the photogenerated free carriers, $\tau$ is the time constant for first-order recombination processes and $B$ is the second-order recombination coefficient. The general solution to Eq. 2 is

$$n = \frac{C}{\exp(At) - CB/A}, \qquad (3)$$

where $A = \alpha_{pc} q^2 + \frac{1}{\tau}$, and $C$ is a fitting parameter related to the initial photogenerated free charge carrier density. When the second-order recombination process is negligible, the solution simplifies to an exponential decay $n = C\exp(-At)$ due to first-order recombination and carrier diffusion. The recombination and carrier diffusion processes can be distinguished due to their different dependences on the grating wavevector $q$, which can be conveniently varied in the experiment by changing the grating period[27].

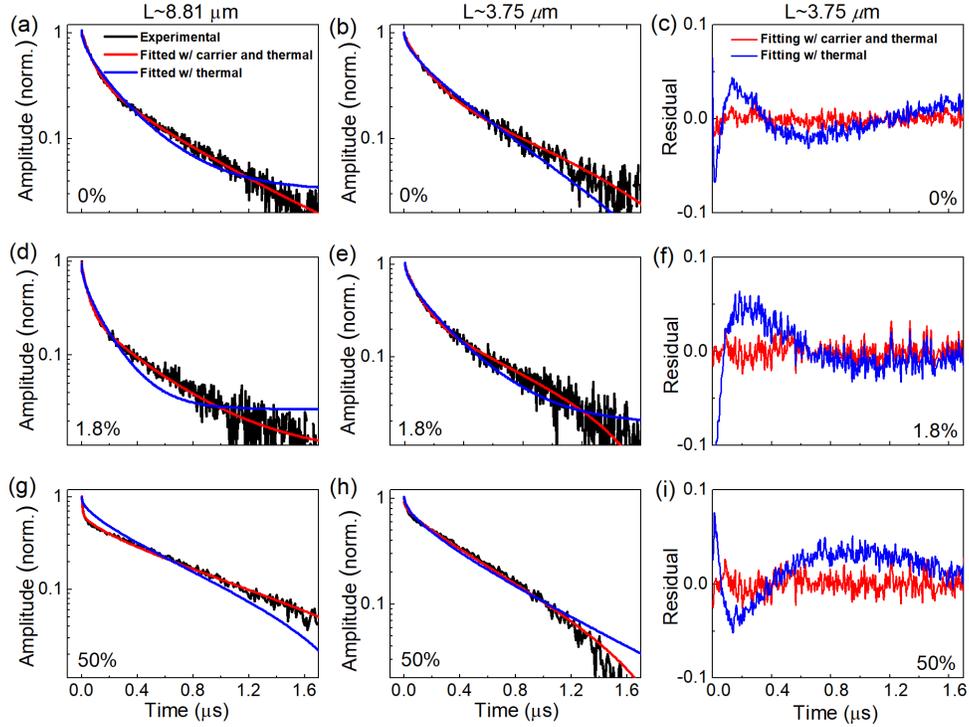

**Figure 2.** The measured decay curves of the transient grating magnitudes and fitting curves for (a-c) neat P3HT, (d-f) 1.8% blend and (g-i) 50% blend. Red solid lines represent fittings performed with combined carrier and thermal contributions, while the blue ones are with thermal contribution only. The left and middle panels are for the grating period of 8.81 and 3.75 μm, respectively, while the right panels are for the residuals of corresponding fittings for 3.75 μm grating data.

Taking into account the contributions from thermal transport and carrier dynamics, we combined Eqs. 1 and 3 to fit the experimental results, as shown in Fig. 2. The fitting residuals are shown in Fig. 2(c)(f)(i), clearly indicating the significant improvement of the fits when both thermal and carrier contributions are included. For neat P3HT and 1.8% blend samples, we found that the experimental results can be well fitted with only thermal transport and first-order recombination processes, while the experimental results for the 50% blend samples can only be fitted if the second-order recombination process is also included. This result reflects that the concentration of free charge carriers in neat P3HT and the 1.8% blend is low due to the low density of donor/acceptor interfaces to generate free charge carriers, so that first-order recombination processes, such as trap-assisted recombination, dominate. We further verified the absence of higher-order recombination processes in these samples by varying the optical power of the pump and the probe beams (see Supplementary Information), where the decay rate $A$ was found to be independent of the optical power (and thus the free charge carrier concentration). In contrast, in the 50% blend samples, which are employed for OPV fabrication, the interpenetration networks of the blend are known to be optimized and the interfacial area is maximized, resulting in optimal exciton dissociation at donor/acceptor interfaces[36,37]. Therefore, the free carrier concentration in the 50% blend samples is much higher and the second-order bimolecular recombination among free charge carriers becomes appreciable.

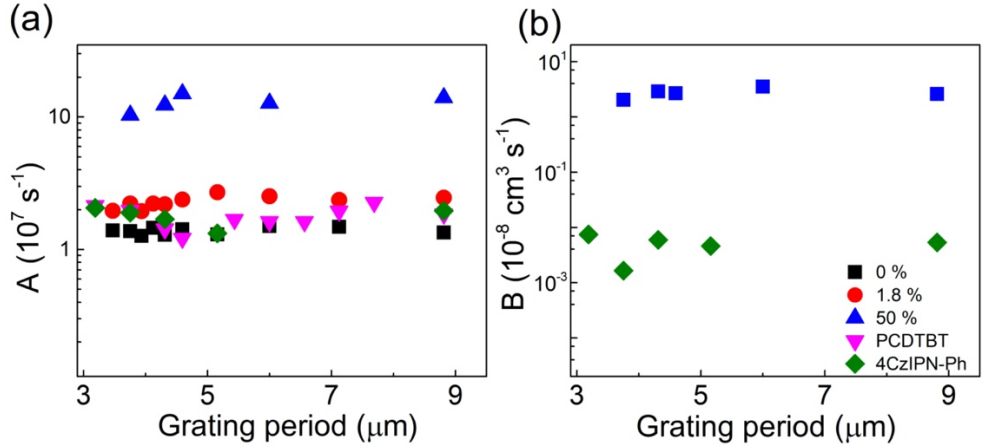

**Figure 3.** The fitted carrier decay rates as a function of grating periods and blend weight ratios: (a) First-order rates for neat P3HT, 1.8% and 50% P3HT:PCBM blends, neat PCDTBT, and 4CzIPN-Ph; (b) Second-order rate for the 50% P3HT:PCBM blend and 4CzIPN-Ph. The fitting uncertainty of A is ~5% for neat P3HT and 1.8% blend and ~10% for 50% blend. The average fitting uncertainty of B is ~10% for the 50% blend and ~18% for 4CzIPN-Ph.

The first-order decay rate $A$ and the second-order recombination coefficient $B$ extracted from our measurements as a function of the grating period are shown in Fig. 3. It is observed that there is no apparent dependence of the first-order decay rate $A$ on the grating period. This indicates that the carrier diffusivity is below the detection limit of the TG technique with micrometer-scale grating periods. Higher resolution of the diffusion coefficients can potentially be obtained with smaller grating periods [16,31]. Given the number of fitting parameters (three for neat P3HT and 1.8% blends, four for 50% blends), the fitting sensitivity and uncertainty of these fitting parameters are critical. We analyzed the fitting uncertainty of each parameter by varying the parameter from the best fitted value until the accumulated deviation of the fitted curve from the experimental curve reaches 10%. The fitting uncertainty of the first-order decay rate $A$ obtained this way is on the order of 5% in neat P3HT and 1.8% blends, and roughly 10% in 50% blend (not shown in the Fig. 3a for clarity). The fitting uncertainty of the second-order recombination coefficient $B$ is roughly 10% for the 50% blend. The first-order decay rates $A$ are $1.4 \times 10^7$ s$^{-1}$, $2.4 \times 10^7$ s$^{-1}$, and $1.3 \times 10^8$ s$^{-1}$ for neat P3HT, 1.8% blend and 50% blend, respectively. The increase in the first-order rate upon increasing PCBM loading is consistent with the blend morphological change, with more free charge carriers generated close to heterojunctions for trapping. Using time-resolved microwave conductivity (TRMC) spectroscopy, Ferguson et al. reported[9] a first-order decay rate of $4.9\times10^7$ s$^{-1}$ that was attributed to trapping of mobile holes in P3HT that was independent of the PCBM concentration.

To generalize the applicability of the TG spectroscopy to other organic semiconducting materials, we also studied a neat film of poly[N-9"-hepta-decanyl-2,7-carbazole-alt-5,5-(4′,7′-di-2-thienyl-2′,1′,3′-benzothiadiazole) (PCDTBT), another typical donor material used in OPVs[38], and a film of 2,4,5,6-tetrakis(3,6-diphenylcarbazol-9-yl)isophthalonitrile (4CzIPN-Ph), a molecular fluorescent material used in organic light emitting devices[39]. The TG decay curves are given in the Supplementary Information and were analyzed with the same process. The first-order decay rates *A* as a function of the grating period is given in Fig. 3a. The first-order decay rate is $\sim1.2\times10^7$ s$^{-1}$ for PCDTBT and $\sim1.7\times10^7$ s$^{-1}$ for 4CzIPN-Ph, which are comparable with neat P3HT.

The second-order recombination rate coefficient in the 50% P3HT:PCBM blends, as shown in Fig. 3b, is $\sim2.7\times10^{-8}$ cm$^3$/s. There are several possible mechanisms for the observed second-order recombination process. The high concentration of free charge carriers in the 50% blend can lead to more frequent annihilation processes between free charge carriers, whose recombination coefficients, however, are typically on the order of $10^{-12}$ cm$^3$/s in the P3HT:PCBM system due to the significantly suppressed Langevin prefactor, as measured by TRMC and transient absorption (TA) in annealed samples[9,40,41]. Other studies based on TRMC and TA[6,8] have reported annihilation processes between excitons and free carriers with a second-order recombination coefficient on the order of $10^{-8}$ cm$^3$/s, but typically observed within a shorter time window (< 1 ns) when a significant amount of excitons remain. In the time window of our TG measurement (tens to hundreds of ns), the exciton density is expected to be very low in the 50% blend, and thus the exciton-carrier recombination is unlikely to be responsible for the second-order process we observed. One possibility is the interaction between free charge carriers with the charge transfer states that extensively exist at the donor/acceptor interfaces before dissociation into free charge carriers[42]. Although there has not been other direct experimental evidence for bimolecular recombination between free charge carriers and charge transfer states, multiscale simulations have indicated the coexistence of bound localized charge transfer states at the donor/acceptor interfaces with a high concentration of delocalized space-separated states[43]. Further studies are needed to confirm this hypothesis and will be part of our future pursuits. We also found an appreciable contribution from bimolecular recombinations in 4CzIPN-Ph, with a rate coefficient of $\sim 5\times 10^{-11}$ cm$^3$/s, most likely due to free charge recombinations.

In summary, we demonstrated the capability of probing photocarrier dynamics and thermal transport simultaneously in a semiconducting polymer thin-film-on-substrate system. Although thermal transport is dominated by the substrate, we were able to separate the contribution from the photocarrier dynamics in the organic thin films. Our study indicates that TG spectroscopy is a promising technique to simultaneously characterize charge carrier transport, recombination and thermal transport in semiconducting polymers and other organic and hybrid optoelectronic materials, provided that thicker samples are used and/or sub-micron grating periods are achieved[16,31].

**Acknowledgements**

This work is based on research supported by a Sony Faculty Innovation Award. T.Q.N. acknowledges support from the Organic Materials Chemistry Program, Air Force Office of Scientific Research (grant no. FA9550-19-1-0348).

**Data Availability**

The data that support the findings of this study are available from the corresponding authors upon reasonable request.

**Reference**

(1) Peplow, M. The Plastics Revolution: How Chemists Are Pushing Polymers to New Limits. *Nature News* **2016**, *536* (7616), 266. https://doi.org/10.1038/536266a.

(2) Kim, J. Y.; Lee, K.; Coates, N. E.; Moses, D.; Nguyen, T.-Q.; Dante, M.; Heeger, A. J. Efficient Tandem Polymer Solar Cells Fabricated by All-Solution Processing. *Science* **2007**, *317* (5835), 222–225. https://doi.org/10.1126/science.1141711.

(3) Yu, G.; Gao, J.; Hummelen, J. C.; Wudl, F.; Heeger, A. J. Polymer Photovoltaic Cells: Enhanced Efficiencies via a Network of Internal Donor-Acceptor Heterojunctions. *Science* **1995**, *270* (5243), 1789–1791. https://doi.org/10.1126/science.270.5243.1789.

(4) Mikhnenko, O. V.; Blom, P. W. M.; Nguyen, T.-Q. Exciton Diffusion in Organic Semiconductors. *Energy & Environmental Science* **2015**, *8* (7), 1867–1888. https://doi.org/10.1039/C5EE00925A.

(5) Proctor, C. M.; Kuik, M.; Nguyen, T.-Q. Charge Carrier Recombination in Organic Solar Cells. *Progress in Polymer Science* **2013**, *38* (12), 1941–1960. https://doi.org/10.1016/j.progpolymsci.2013.08.008.

(6) Hodgkiss, J. M.; Albert-Seifried, S.; Rao, A.; Barker, A. J.; Campbell, A. R.; Marsh, R. A.; Friend, R. H. Exciton-Charge Annihilation in Organic Semiconductor Films. *Advanced Functional Materials* **2012**, *22* (8), 1567–1577. https://doi.org/10.1002/adfm.201102433.

(7) Hwang, I.-W.; Moses, D.; Heeger, A. J. Photoinduced Carrier Generation in P3HT/PCBM Bulk Heterojunction Materials. *J. Phys. Chem. C* **2008**, *112* (11), 4350–4354. https://doi.org/10.1021/jp075565x.

(8) Ferguson, A. J.; Kopidakis, N.; Shaheen, S. E.; Rumbles, G. Quenching of Excitons by Holes in Poly(3-Hexylthiophene) Films. *J. Phys. Chem. C* **2008**, *112* (26), 9865–9871. https://doi.org/10.1021/jp7113412.

(9) Ferguson, A. J.; Kopidakis, N.; Shaheen, S. E.; Rumbles, G. Dark Carriers, Trapping, and Activation Control of Carrier Recombination in Neat P3HT and P3HT:PCBM Blends. *J. Phys. Chem. C* **2011**, *115* (46), 23134–23148. https://doi.org/10.1021/jp208014v.

(10) Mihailetchi, V. D.; Wildeman, J.; Blom, P. W. M. Space-Charge Limited Photocurrent. *Phys. Rev. Lett.* **2005**, *94* (12), 126602. https://doi.org/10.1103/PhysRevLett.94.126602.

(11) Jørgensen, M.; Norrman, K.; Krebs, F. C. Stability/Degradation of Polymer Solar Cells. *Solar Energy Materials and Solar Cells* **2008**, *92* (7), 686–714. https://doi.org/10.1016/j.solmat.2008.01.005.

(12) Wang, X.; Ho, V.; Segalman, R. A.; Cahill, D. G. Thermal Conductivity of High-Modulus Polymer Fibers. *Macromolecules* **2013**, *46* (12), 4937–4943. https://doi.org/10.1021/ma400612y.

(13) Weathers, A.; Khan, Z. U.; Brooke, R.; Evans, D.; Pettes, M. T.; Andreasen, J. W.; Crispin, X.; Shi, L. Significant Electronic Thermal Transport in the Conducting Polymer Poly(3,4-Ethylenedioxythiophene). *Advanced Materials* **2015**, *27* (12), 2101–2106. https://doi.org/10.1002/adma.201404738.

(14) Xu, Y.; Kraemer, D.; Song, B.; Jiang, Z.; Zhou, J.; Loomis, J.; Wang, J.; Li, M.; Ghasemi, H.; Huang, X.; Li, X.; Chen, G. Nanostructured Polymer Films with Metal-like Thermal Conductivity. *arXiv:1708.06416* **2017**.

(15) Johnson, J. A.; Maznev, A. A.; Bulsara, M. T.; Fitzgerald, E. A.; Harman, T. C.; Calawa, S.; Vineis, C. J.; Turner, G.; Nelson, K. A. Phase-Controlled, Heterodyne Laser-Induced

Transient Grating Measurements of Thermal Transport Properties in Opaque Material. *Journal of Applied Physics* **2012**, *111* (2), 023503. https://doi.org/10.1063/1.3675467.
(16) Robbins, A. B.; Drakopoulos, S. X.; Martin-Fabiani, I.; Ronca, S.; Minnich, A. J. Ballistic Thermal Phonons Traversing Nanocrystalline Domains in Oriented Polyethylene. *PNAS* **2019**, *116* (35), 17163–17168. https://doi.org/10.1073/pnas.1905492116.
(17) Käding, O. W.; Skurk, H.; Maznev, A. A.; Matthias, E. Transient Thermal Gratings at Surfaces for Thermal Characterization of Bulk Materials and Thin Films. *Appl. Phys. A* **1995**, *61* (3), 253–261. https://doi.org/10.1007/BF01538190.
(18) Vega-Flick, A.; Duncan, R. A.; Eliason, J. K.; Cuffe, J.; Johnson, J. A.; Peraud, J.-P. M.; Zeng, L.; Lu, Z.; Maznev, A. A.; Wang, E. N.; Alvarado-Gil, J. J.; Sledzinska, M.; Sotomayor Torres, C. M.; Chen, G.; Nelson, K. A. Thermal Transport in Suspended Silicon Membranes Measured by Laser-Induced Transient Gratings. *AIP Advances* **2016**, *6* (12), 121903. https://doi.org/10.1063/1.4968610.
(19) Vega-Flick, A.; Jung, D.; Yue, S.; Bowers, J. E.; Liao, B. Reduced Thermal Conductivity of Epitaxial GaAs on Si Due to Symmetry-Breaking Biaxial Strain. *Phys. Rev. Materials* **2019**, *3* (3), 034603. https://doi.org/10.1103/PhysRevMaterials.3.034603.
(20) Cowan, S. R.; Roy, A.; Heeger, A. J. Recombination in Polymer-Fullerene Bulk Heterojunction Solar Cells. *Phys. Rev. B* **2010**, *82* (24), 245207. https://doi.org/10.1103/PhysRevB.82.245207.
(21) Wang, H.; Wang, H.-Y.; Gao, B.-R.; Wang, L.; Yang, Z.-Y.; Du, X.-B.; Chen, Q.-D.; Song, J.-F.; Sun, H.-B. Exciton Diffusion and Charge Transfer Dynamics in Nano Phase-Separated P3HT/PCBM Blend Films. *Nanoscale* **2011**, *3* (5), 2280–2285. https://doi.org/10.1039/C0NR01002B.
(22) Mikhnenko, O. V.; Kuik, M.; Lin, J.; Kaap, N. van der; Nguyen, T.-Q.; Blom, P. W. M. Trap-Limited Exciton Diffusion in Organic Semiconductors. *Advanced Materials* **2014**, *26* (12), 1912–1917. https://doi.org/10.1002/adma.201304162.
(23) Feng, X.; Wang, X. Thermophysical Properties of Free-Standing Micrometer-Thick Poly(3-Hexylthiophene) Films. Thin Solid Films **2011**, *519* (16), 5700–5705. https://doi.org/10.1016/j.tsf.2011.03.043.
(24) Duda, J. C.; Hopkins, P. E.; Shen, Y.; Gupta, M. C. Thermal Transport in Organic Semiconducting Polymers. *Appl. Phys. Lett.* **2013**, *102* (25), 251912. https://doi.org/10.1063/1.4812234.
(25) Xu, Y.; Wang, X.; Zhou, J.; Song, B.; Jiang, Z.; Lee, E. M. Y.; Huberman, S.; Gleason, K. K.; Chen, G. Molecular Engineered Conjugated Polymer with High Thermal Conductivity. *Science Advances* **2018**, *4* (3), eaar3031. https://doi.org/10.1126/sciadv.aar3031.
(26) Goodno, G. D.; Dadusc, G.; Miller, R. J. D. Ultrafast Heterodyne-Detected Transient-Grating Spectroscopy Using Diffractive Optics. *J. Opt. Soc. Am. B* **1998**, *15* (6), 1791–1794. https://doi.org/10.1364/JOSAB.15.001791.
(27) Mahmood, F.; Alpichshev, Z.; Lee, Y.-H.; Kong, J.; Gedik, N. Observation of Exciton–Exciton Interaction Mediated Valley Depolarization in Monolayer $MoSe_2$. *Nano Lett.* **2018**, *18* (1), 223–228. https://doi.org/10.1021/acs.nanolett.7b03953.
(28) Kim, T.; Ding, D.; Yim, J.-H.; Jho, Y.-D.; Minnich, A. J. Elastic and Thermal Properties of Free-Standing Molybdenum Disulfide Membranes Measured Using Ultrafast Transient Grating Spectroscopy. *APL Materials* **2017**, *5* (8), 086105. https://doi.org/10.1063/1.4999225.

(29) Sjodin, T.; Petek, H.; Dai, H.-L. Ultrafast Carrier Dynamics in Silicon: A Two-Color Transient Reflection Grating Study on a (111) Surface. *Phys. Rev. Lett.* **1998**, *81* (25), 5664–5667. https://doi.org/10.1103/PhysRevLett.81.5664.

(30) Johnson, J. A.; Maznev, A. A.; Cuffe, J.; Eliason, J. K.; Minnich, A. J.; Kehoe, T.; Torres, C. M. S.; Chen, G.; Nelson, K. A. Direct Measurement of Room-Temperature Nondiffusive Thermal Transport Over Micron Distances in a Silicon Membrane. *Phys. Rev. Lett.* **2013**, *110* (2), 025901. https://doi.org/10.1103/PhysRevLett.110.025901.

(31) Bencivenga, F.; Mincigrucci, R.; Capotondi, F.; Foglia, L.; Naumenko, D.; Maznev, A. A.; Pedersoli, E.; Simoncig, A.; Caporaletti, F.; Chiloyan, V.; Cucini, R.; Dallari, F.; Duncan, R. A.; Frazer, T. D.; Gaio, G.; Gessini, A.; Giannessi, L.; Huberman, S.; Kapteyn, H.; Knobloch, J.; Kurdi, G.; Mahne, N.; Manfredda, M.; Martinelli, A.; Murnane, M.; Principi, E.; Raimondi, L.; Spampinati, S.; Spezzani, C.; Trovò, M.; Zangrando, M.; Chen, G.; Monaco, G.; Nelson, K. A.; Masciovecchio, C. Nanoscale Transient Gratings Excited and Probed by Extreme Ultraviolet Femtosecond Pulses. *Science Advances* **2019**, *5* (7), eaaw5805. https://doi.org/10.1126/sciadv.aaw5805.

(32) Vollbrecht, J.; Brus, V. V.; Ko, S.-J.; Lee, J.; Karki, A.; Cao, D. X.; Cho, K.; Bazan, G. C.; Nguyen, T.-Q. Quantifying the Nongeminate Recombination Dynamics in Nonfullerene Bulk Heterojunction Organic Solar Cells. *Advanced Energy Materials* **2019**, *9* (32), 1901438. https://doi.org/10.1002/aenm.201901438.

(33) Piris, J.; Dykstra, T. E.; Bakulin, A. A.; Loosdrecht, P. H. M. van; Knulst, W.; Trinh, M. T.; Schins, J. M.; Siebbeles, L. D. A. Photogeneration and Ultrafast Dynamics of Excitons and Charges in P3HT/PCBM Blends. *J. Phys. Chem. C* **2009**, *113* (32), 14500–14506. https://doi.org/10.1021/jp904229q.

(34) Sheng, C.-X.; Tong, M.; Singh, S.; Vardeny, Z. V. Experimental Determination of the Charge/Neutral Branching Ratio η in the Photoexcitation of π-Conjugated Polymers by Broadband Ultrafast Spectroscopy. *Phys. Rev. B* **2007**, *75* (8), 085206. https://doi.org/10.1103/PhysRevB.75.085206.

(35) Brus, V. V.; Proctor, C. M.; Ran, N. A.; Nguyen, T.-Q. Capacitance Spectroscopy for Quantifying Recombination Losses in Nonfullerene Small-Molecule Bulk Heterojunction Solar Cells. *Advanced Energy Materials* **2016**, *6* (11), 1502250. https://doi.org/10.1002/aenm.201502250.

(36) Lee, C.-K.; Pao, C.-W.; Chu, C.-W. Multiscale Molecular Simulations of the Nanoscale Morphologies of P3HT:PCBM Blends for Bulk Heterojunction Organic Photovoltaic Cells. *Energy Environ. Sci.* **2011**, *4* (10), 4124–4132. https://doi.org/10.1039/C1EE01508G.

(37) Liang, Y.; Xu, Z.; Xia, J.; Tsai, S.-T.; Wu, Y.; Li, G.; Ray, C.; Yu, L. For the Bright Future-Bulk Heterojunction Polymer Solar Cells with Power Conversion Efficiency of 7.4%. *Advanced Materials* **2010**, *22* (20), E135–E138. https://doi.org/10.1002/adma.200903528.

(38) Park, S. H.; Roy, A.; Beaupré, S.; Cho, S.; Coates, N.; Moon, J. S.; Moses, D.; Leclerc, M.; Lee, K.; Heeger, A. J. Bulk Heterojunction Solar Cells with Internal Quantum Efficiency Approaching 100%. *Nature Photonics* **2009**, *3* (5), 297–302. https://doi.org/10.1038/nphoton.2009.69.

(39) Yurash, B.; Nakanotani, H.; Olivier, Y.; Beljonne, D.; Adachi, C.; Nguyen, T.-Q. Photoluminescence Quenching Probes Spin Conversion and Exciton Dynamics in Thermally Activated Delayed Fluorescence Materials. *Advanced Materials* **2019**, *31* (21), 1804490. https://doi.org/10.1002/adma.201804490.

(40) Murthy, D. H. K.; Melianas, A.; Tang, Z.; Juška, G.; Arlauskas, K.; Zhang, F.; Siebbeles, L. D. A.; Inganäs, O.; Savenije, T. J. Origin of Reduced Bimolecular Recombination in Blends of Conjugated Polymers and Fullerenes. *Advanced Functional Materials* **2013**, *23* (34), 4262–4268. https://doi.org/10.1002/adfm.201203852.

(41) Clarke, T. M.; Jamieson, F. C.; Durrant, J. R. Transient Absorption Studies of Bimolecular Recombination Dynamics in Polythiophene/Fullerene Blend Films. *J. Phys. Chem. C* **2009**, *113* (49), 20934–20941. https://doi.org/10.1021/jp909442s.

(42) Coropceanu, V.; Chen, X.-K.; Wang, T.; Zheng, Z.; Brédas, J.-L. Charge-Transfer Electronic States in Organic Solar Cells. *Nature Reviews Materials* **2019**, *4* (11), 689–707. https://doi.org/10.1038/s41578-019-0137-9.

(43) D'Avino, G.; Muccioli, L.; Olivier, Y.; Beljonne, D. Charge Separation and Recombination at Polymer–Fullerene Heterojunctions: Delocalization and Hybridization Effects. *J. Phys. Chem. Lett.* **2016**, *7* (3), 536–540. https://doi.org/10.1021/acs.jpclett.5b02680.

**Supplementary Information: Transient Grating Spectroscopy of Photocarrier Dynamics in Semiconducting Polymer Thin Films**

**Wenkai Ouyang[a], Yu Li[a], Brett Yurash[b], Nora Schopp[b], Alejandro Vega-Flick[a], Viktor Brus[bc], Thuc-Quyen Nguyen[b] and Bolin Liao[a]**

[a]Department of Mechanical Engineering, University of California, Santa Barbara, 93106

[b]Department of Chemistry and Biochemistry, University of California, Santa Barbara, 93106

[c]Department of Physics, School of Sciences and Humanities, Nazarbayev University, Nur-Sultan City, 010000, Republic of Kazakhstan

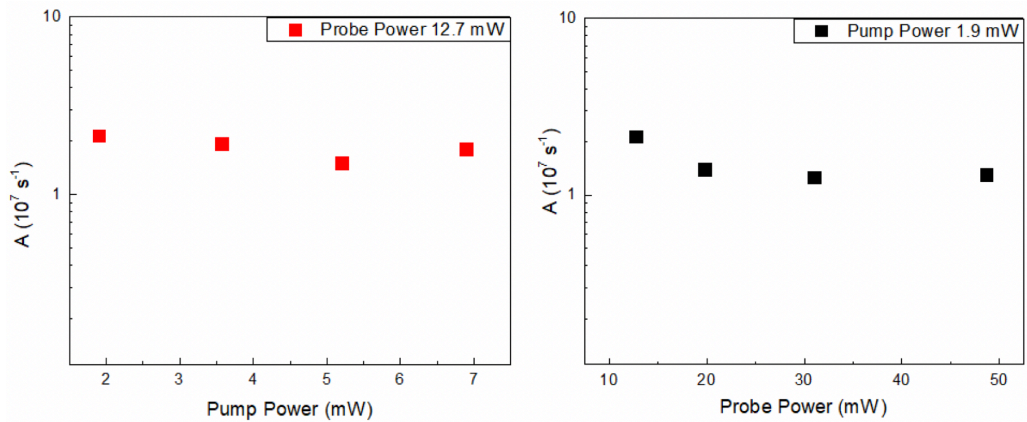

**Figure S1.** The dependence of the first-order decay rate of the neat P3HT sample on the optical powers of the pump and the probe beam. The decay rate in neat P3HT is independent of the optical powers, confirming the first-order nature of the free carrier recombination process.

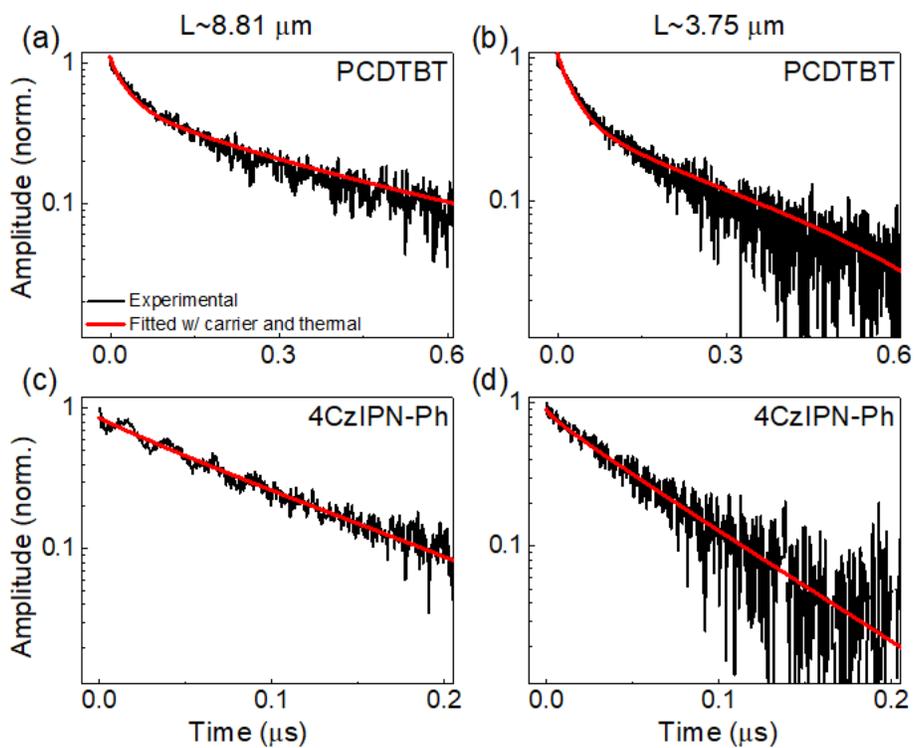

**Figure S2.** The grating decays and fittings for PCDTBT and 4CzIPN-Ph for two grating periods (a,c) 8.81 μm and (b,d) 3.75 μm. The black lines are from the experimental data, and the red lines are the fittings considering the carrier and thermal decays simultaneously. The oscillations in Fig. 2c are due to acoustic waves in the air above the sample surface generated by the pump pulse.